\documentclass[12pt]{article}
\usepackage{amsmath,amsfonts, amssymb,amsthm}
\usepackage[dvipsnames]{xcolor}
\usepackage[latin1]{inputenc}
\usepackage{enumitem}
\usepackage{authblk}
\usepackage{subfigure}
\usepackage{setspace}
\usepackage[ruled]{algorithm2e} 

\usepackage{etoolbox}

\usepackage[hypcap=false]{caption}

\PassOptionsToPackage{hyphens}{url}
\usepackage{hyperref}

\hypersetup{
     colorlinks=true,
     linkcolor=blue,
     filecolor=blue,
     citecolor = blue,      
     urlcolor=cyan,
     }

\usepackage{natbib}

\usepackage[margin=1in]{geometry}
\usepackage[normalem]{ulem}

\usepackage{graphicx}
\usepackage{color}

\newtheorem{theorem}{Theorem}
\newtheorem{lemma}{Lemma}
\newtheorem{proposition}{Proposition}

\newtheorem{assumption}{Assumption}
\newtheorem{definition}{Definition} 

\newtoggle{comments}
\toggletrue{comments}

\iftoggle{comments}{
\newcommand{\at}[1]{{\color{orange} AT: #1}}
\newcommand{\sv}[1]{{\color{red}  SV: #1}}
\newcommand{\tn}[1]{{\color{purple} TN:  #1}}

}{
\newcommand{\at}[1]{}
\newcommand{\sv}[1]{}
\newcommand{\tn}[1]{}
}

\newcommand{\proofst}{\proof{Proof. } }
\newcommand{\proofen}{\hfill\endproof}

\newcommand{\sample}{\mathcal{S}}
\newcommand{\supply}[1]{c^{#1}}

\newcommand{\epss}{\epsilon_s}
\newcommand{\epsf}{\epsilon_f}

\newcommand{\expect}[1]{\mathbb{E}\left[#1\right]}

\newcommand{\hoef}{x} 
\newcommand{\hoefset}{\mathcal{X}} 
\newcommand{\hoefrand}{X}

\newcommand{\avgpref}{\mu}
\newcommand{\numtypes}{\tau}

\newcommand{\agent}{i} 

\newcommand{\altagent}{\ell} 
\newcommand{\good}{j}
\newcommand{\ngoods}{m} 
\newcommand{\agentset}{N}

\newcommand{\nagents}{n}
\newcommand{\goodsset}{M}
\newcommand{\goodset}{\goodsset}
\newcommand{\Nn}{\nagents}
\newcommand{\knum}{k}

\newcommand{\typeset}{\Theta}
\newcommand{\sampletypeset}{\typeset_{\sample}}

\newcommand{\iter}{i}

\newcommand{\optsub}[2]{{\hat{#1}}_{#2}}
\newcommand{\optsup}[2]{{\hat{#1}}^{#2}}
\newcommand{\optboth}[3]{{\hat{#1}}_{#2}^{#3}}

\newcommand{\nsample}{s}

\newcommand{\prices}{{\bf p}}
\newcommand{\price}[1]{p^{#1}} 
\newcommand{\alloc}[2]{x_{#2}^{#1}} 
\newcommand{\allocs}{\mathbf{X}}
\newcommand{\rvalloc}[2]{{\mathcal X}_{#2}^{#1}} 

\newcommand{\alloct}{{\bf x}}
\newcommand{\ralloc}[2]{y_{#2}^{#1}} 
\newcommand{\rallocv}[1]{{\bf y}_{#1}} 
\newcommand{\rvralloc}[2]{{\mathcal Y}_{#2}^{#1}} 
\newcommand{\rvrallocv}[1]{{\mathcal Y}_{#1}} 
\newcommand{\allocv}[1]{{\bf x}_{#1}}
\newcommand{\rvallocv}[1]{{\mathcal X}_{#1}}
\newcommand{\gallocv}[1]{{\bf x}_{#1}}
\newcommand{\optprice}[1]{\optsup{p}{#1}}
\newcommand{\optprices}{\optsup{\prices}{}}
\newcommand{\optbudget}[1]{\optsub{b}{#1}}
\newcommand{\optbudgets}{\optsub{\budgets}{}}
\newcommand{\optsupply}[1]{\optsup{c}{#1}}
\newcommand{\optsupplyvector}{\optsup{\supplyvector}{}}
\newcommand{\optallocs}{\optsub{\allocs}{}}
\newcommand{\optgallocv}[1]{\optsub{\alloct}{#1}}
\newcommand{\optgalloc}[2]{\optboth{x}{#2}{#1}}

\newcommand{\kn}{k}

\newcommand{\ballocv}[1]{{\bf w}_{#1}}
\newcommand{\balloc}[2]{w_{#1}^{#2}} 
\newcommand{\bundle}{{\bf x}}

\newcommand{\ignore}[1]{}

\newcommand{\epsnn}{\sampleeps \nagents}
\newcommand{\budgeteps}{\epsilon_{\budget{}}}
\newcommand{\approxeps}{\epsf}
\newcommand{\sampleeps}{\epsilon_{\nagents}}
\newcommand{\epsn}{\sampleeps}
\newcommand{\epsHoef}{\epsilon}
\newcommand{\epsBern}{\epsilon}

\newcommand{\choices}{\Psi}
\newcommand{\budget}[1]{b_{#1}}

\newcommand{\ttype}{\theta}
\newcommand{\tempy}[1]{\eta^\good_{#1}(\triple)}

\newcommand{\mech}{DACEEI}

\newcommand{\outputtypeset}{\Lambda}

\newcommand{\triple}{q}
\newcommand{\mechanism}{\Phi}
\newcommand{\epsilons}{\alleps>0}

\newcommand{\epschernoff}{\epsilon}

\newcommand{\epsft}{ \frac{\epsf}{4}}

\newcommand{\epsfa}{ \frac{\epsf}{4}}
\newcommand{\epsfb}{ \frac{\epsf}{2}}
\newcommand{\epsfc}{ \frac{3\epsf}{4}}
\newcommand{\epsfd}{ \epsf}
\newcommand{\repsa}{\frac{1}{2\nagents}}
\newcommand{\repsb}{\frac{1}{3\nagents}}
\newcommand{\repsc}{\frac{1}{6\nagents}}

\newcommand{\eventone}{\mathbb{I}_k(\triple, \good)}
\newcommand{\eventtwo}{\mathbb{I}_n(\triple, \good)}

\newcommand{\eventy}{\mathbb{I}_y(\triple,\good)}

\newcommand{\randombudget}[1]{\mathcal{B}_{#1}}
\newcommand{\optrandombudget}[1]{\optsub{\mathcal{B}}{#1}}

\newcommand{\optbundle}[1]{\optsub{\mathcal{X}}{#1}}
\newcommand{\optbundlev}{\optsub{\mathcal{X}}{}}
\newcommand{\budgetfunc}{f}

\newcommand{\bundlev}{\mathcal{X}}

\newcommand{\budgets}{\mathbf{b}}
\newcommand{\emptybundle}{{\bf 0}}
\newcommand{\util}{u}

\newcommand{\y}{{\bf y}}
\newcommand{\x}{{\bf x}}
\newcommand{\e}{{\bf e}}
\newcommand{\ooo}{{\bf 0}}

\newcommand{\supplyvector}{\mathbf{\supply{}}}

\newcommand{\econ}{E}
\newcommand{\ceei}{CEEI}
\newcommand{\eceei}{ECEEI}
\newcommand{\aceei}{ACEEI}

\newcommand{\alleps}{\budgeteps, \sampleeps,\approxeps}

\newcommand{\epsb}{\budgeteps}
\newcommand{\gsset}{G}
\newcommand{\daisyeps}{(\alleps)}

\newcommand{\agentdef}{k}
\newcommand{\iden}{\pi^I}
\newcommand{\ocam}{OCAM}
\newcommand{\OCAM}{OCAM}

\newcommand{\realsample}{\epss \nagents}

\newcommand{\ocamlong}{online combinatorial assignment mechanism}

\makeatletter
\g@addto@macro{\definition}{\upshape}
\makeatother

\title{Dynamic Combinatorial Assignment}
\author{
Th\`{a}nh Nguyen\thanks{Daniels School of Management, Purdue University, 403 W. State Street, West Lafayette, Indiana, 47906, United States.  E-mail: {\tt nguye161@purdue.edu}.}
, Alexander Teytelboym\thanks{Department of Economics and St.~Catherine's College, University of Oxford, Oxford, OX1 3UQ, United Kingdom. Email: {\tt alexander.teytelboym@economics.ox.ac.uk}.}
, Shai Vardi\thanks{Daniels School of Management, Purdue University, 403 W. State Street, West Lafayette, Indiana, 47906, United States.  E-mail: {\tt svardi@purdue.edu}.}
}
\date{\today}
\begin{document}

\maketitle

\begin{abstract}

We study a model of dynamic combinatorial assignment of indivisible objects without money. We introduce a new solution concept called \emph{dynamic approximate competitive equilibrium from equal incomes} (DACEEI), which stipulates that markets must approximately clear in almost all time periods. A naive repeated application of approximate competitive equilibrium from equal incomes \citep{budish2011combinatorial} does not yield a desirable outcome because the approximation error in market-clearing compounds quickly over time. We therefore develop a new version of the static approximate competitive equilibrium from carefully constructed random budgets which ensures that, in expectation, markets clear exactly. We then use it to design the online combinatorial assignment mechanism (OCAM) which implements a DACEEI with high probability. The OCAM is (i) group-strategyproof up to one object (ii) envy-free up to one object for almost all agents (iii) approximately market-clearing in almost all periods with high probability when the market is large and arrivals are random. Applications include refugee resettlement, daycare assignment, and airport slot allocation.

\end{abstract}

\maketitle
\doublespacing

\ignore{
Daniela Saban\\
Sid Banerjee\\
Daniel Freund\\
Eduardo Azevedo\\
Eric Budish\\
}

\section{Introduction}

In many complex allocation  problems, such as daycare assignments, refugee resettlement and airport slot allocation, the market designer needs to assign combinations of objects to agents, but cannot use money. In such \emph{combinatorial assignment} settings, the only efficient and strategyproof mechanisms are dictatorships which can be grossly unfair (e.g., \citealt{klaus2002strategy}). One approach that circumvents these stark tradeoffs is to use \emph{pseudomarkets}: agents are allocated (equal) budgets of tokens and can use them to trade the resources. An allocation in a \emph{competitive equilibrium from equal incomes} (CEEI) in this setting is efficient and eliminates envy \citep{varian1974equity}. Moreover, in large markets truthfully revealing preferences becomes a dominant strategy in a mechanism that implements a CEEI \citep{azevedo2019strategy}. 

Unfortunately, when resources are indivisible and/or highly complementary, a CEEI might not necessarily exist \citep{hylland1979efficient}.
A seminal paper by \citet{budish2011combinatorial} offered an alternative to the (exact) CEEI called \emph{approximate} CEEI (\aceei). In an \aceei, agents' budgets are slightly perturbed and the market for each resource clears only approximately. The approximation bound on the excess demand reflects the number of objects that have been left unallocated or need to be added in order to establish an exact CEEI. Thus, an \aceei\ remains nearly envy-free and efficient when the excess demand is small relative to the total supply.

This paper considers pseudomarkets in the context of \emph{dynamic} (online)  
combinatorial assignment problems without monetary transfers. 
The following  examples illustrate our setting.

\begin{enumerate}
    \item Refugee resettlement~\citep[e.g.,][]{ahani2021dynamic}: refugees arrive  over the course of some time. As soon as a refugee family arrives, it needs to be matched to a location. Refugees have preferences over different possible locations for initial resettlement. Refugee families are of different sizes and might require a different amount of services offered at different locations (e.g., school places). 
    \item Daycare allocation~\citep[e.g.,][]{kennes2014day}: 
    families with children demand childcare. 
    Families have preferences over different daycare centers (or over combinations of daycare centers and days of the week). 
    Daycare centers cannot easily add extra staff or rooms so their capacity constraints need to be respected throughout the year.
    
    
    \item Assignment of airport take-off and landing slots~\citep[e.g.,][]{schummer2013assignment}: 
    airlines demand landing slots, gates and taking-off slots for  their flights. 
    The assignment frequently needs to be negotiated hour-by-hour, often when aircraft are still in the air, because of varying weather conditions. Airlines might prefer to have take-off and landing slots that are close to each other to avoid idling the aircraft at the gate. Airports have strict capacities at different gates and for the number of aircraft that can land  or take-off from a runway in any time period which must be respected for safety reasons.
\end{enumerate}


There are two significant hurdles that need to be overcome in order to extend the \aceei\ to dynamic settings. The first  involves making irreversible allocation decisions as agents arrive. As a result, naive repeated use of ACEEI becomes increasingly inefficient because the bound on excess demand scales up linearly with each iteration, making the inefficiency directly proportional to the number of agents in online settings. The second hurdle is the lack of available preference data before agents arrive, which requires the development of an online allocation mechanism that can function without any prior assumptions about the preference distribution.  


\ignore{
arises when dealing with indivisible goods, as CEEI may not exist. One possible solution could be to repeatedly use \aceei. However, this approach may lead to poor results, since the bound on excess demand scales up at the rate of the number of times \aceei\ is used, which can increase linearly with the number of agents in online settings.

To address this, we utilize the  random arrival framework \citep[e.g.,][]{Freeman83}, in which agents arrive in a random order  from an unknown distribution. An efficient mechanism in this model will need to learn preferences over time while allocating resources to them. 
}

We propose a pseudomarket mechanism for dynamic combinatorial assignment problems such as those described above.
Despite the above challenges, the allocation produced by our mechanism satisfies a novel and strong solution concept called \emph{dynamic} \aceei\ (\mech)\footnote{Pronounced `Daisy' for convenience.}, in which markets clear approximately \emph{in every period} (i.e., whenever a new agent or a batch of agents appear), except for a small number of initial periods.  We show that the \mech\ inherits a number of desirable properties of its static counterpart: it is envy-free up to one object (Proposition~\ref{prop:EF}) and ex-post Pareto efficient with respect to realized supply  (Proposition~\ref{prop:PE}). We then describe an online mechanism that implements a \mech\ with high probability while giving almost all agents a limited incentive to misreport their preferences. Importantly, we leverage the random arrival framework \citep[e.g.,][]{Freeman83}, which allows agents to arrive randomly from an \emph{unknown} distribution. This framework necessitates our mechanism to learn preferences over time while effectively allocating resources. We obtain a efficient dynamic combinatorial assignment mechanism in two key steps.

First, in order to control the compounding error in excess demand that would arise from the repeated use of \aceei, we introduce a  new equilibrium concept called \emph{expected} CEEI (\eceei) for static combinatorial assignment settings. 
Given any $\epsilon>0$, an \eceei\  is characterized by  equilibrium prices and distributions over perturbed agent incomes  such that the agents' average consumption at the equilibrium prices satisfies the \emph{exact} market-clearing condition, and the incomes are all in the interval $[1-\epsilon, 1]$. Theorem~\ref{theo:ECEEI} shows that an \eceei\ exists for any $\epsilon>0$. 

Second, we show how to adapt  \eceei\ to dynamic settings to ensure that markets clear approximately in every period, after an initial sample. Specifically, our \ocamlong\ (\ocam) 
has two phases:
\begin{enumerate}
\item For an initial set of arriving agents (which we refer to as the \emph{sample}), allocate the objects according to a (random) serial dictatorship (SD). 
We use the  reported preferences of these agents to calculate an \eceei.  
\item For the remaining agents, allocate each agent their favorite bundle using prices from the sample \eceei\ conditional on the \emph{random} budget determined by their type (i.e., their preference ordering over objects). 
\end{enumerate}

Theorem~\ref{thm1} shows that the \ocam \ is  group-strategyproof up to one object,
envy-free up to one object for all agents outside the sample,  efficient (i.e., market-clearing) with high probability when capacities are sufficiently large and agents arrive in random order, and
 uses up the capacity of each good at a constant rate.
The high-level idea behind the proof is the following: the SD guarantees that the reports of the agents are truthful, and the budgets of the agents outside the  sample are perturbed based on their reported type to maintain incentives for truthtelling.
The \eceei\ ensures that market-clearing conditions (for the sample) are not violated in expectation. Suitable  concentration inequalities are then used to extend the guarantees on the expectation for the sample to bounds on the deviations of the realized allocations in almost every period.


\subsection{Related literature}


Our paper is related to several strands of the literature. 
The first strand is the analysis of equilibrium in static pseudomarkets.
\citet{varian1974equity} offered a striking analysis of the fairness and efficiency properties of the CEEI in a divisible object setting.
\cite{hylland1979efficient} first proposed using pseudomarkets (via trading probability shares) for the allocation of indivisible objects. 
\citet{budish2011combinatorial} dealt with the non-existence of equilibrium in pseudomarkets for combinatorial assignment by introducing the \aceei. Our solution concept, the \mech\, directly extends Budish's \aceei\ to the dynamic setting. 
Our paper is also related to \cite{nguyen2021delta}, which shows the existence of a pseudoequilibrium in an economy with money. 
In this paper, we take our results one step further by demonstrating how to leverage a pseudoequilibrium to construct an \eceei\ using a lottery over random budgets. This approach is a crucial element in extending the combinatorial assignment to dynamic settings.


Second, our paper is related to dynamic pseudomarkets and dynamic fairness problems. Several papers~\citep[e.g.][]{balseiro2019multiagent,gorokh2021monetary} analyze scrip systems; \cite{combe2021dynamic} considers dynamic assignment via spot mechanisms. The key difference in this paper is that we consider fairness and incentive compatibility for each new agent that arrives in every period, whereas the previous models focused on objects arriving over time and long-lived agents.

Another stream of papers  studies the trade-off between efficiency and fairness in the dynamic setting \citep[e.g.][]{Zeng2020,manshadi2021fair}. These papers typically focus on fair division of either divisible or indivisible goods, where the agents have cardinal valuations for the goods, in contrast to our setting where the agents' preferences over bundles of objects are the key feature in our mechanism.

Finally, from a methodological perspective, our paper builds on recent literature on algorithms for online linear programming. Specifically, our paper is most related to \citet{AgrawalOnline} and \citet{MolinaroOnline}, which study a related dynamic setting where agents are associated with the variables of a packing linear program. 
They sample a constant fraction of the agents, solve a dual problem constructed from this sample, and use these prices as the basis of a dynamic binary allocation rule; they use the geometric structure of the solution space to bound the number of possible distinct solutions.
Our setting differs from theirs in several aspects: our input is ordinal, while theirs is cardinal, and their decisions are binary---to allocate or not---while we must allocate a bundle to every agent. Further, our allocations are randomized; our proofs make use of the two different dependence structures of our randomness sources. 
More recently, \citet{Vera} study a related model in which each agent is single-minded, and the decision-maker has to decide whether or not to allocate the bundle of interest to the agent.


\section{Model}
There is a finite set  $\goodsset$ of \emph{goods}, $|\goodsset|=\ngoods$  and a finite set $\agentset$ of agents, $|\agentset| = \nagents$. Each good $\good$ has a finite integer capacity $\supply{\good}\in \mathbb{N}$. 
 There is a permutation $\pi: \agentset \rightarrow \agentset$ on the agents; the permutation defines the order of arrival of the agents. Without loss of generality, agents are indexed by their order of arrival. We allow agents to arrive in batches or one by one. 
 
 A \emph{bundle} $\bundle \in \{0,1\}^m$ contains at most one unit of each good.\footnote{As \citet{budish2011combinatorial} points out, this is without loss of generality as identical units can be simply relabelled as different objects.} We use $(\x-\e^j)^+$  to denote the  bundle $\x$ with 
 object $\good$ removed. 
The bundle consumed by agent $\agent$ is denoted by~$\bundle_{\agent}$. There might be further constraints on consumption (for example, taking off and landing slots that are too close to one another). Let $\choices_\agent \subseteq \{0,1\}^m $ denote the set of \emph{feasible} bundles for agent $\agent$. We assume free disposal, i.e., that $\ooo\in \choices_\agent$ for all $\agent$.
An \emph{allocation} $\allocs=(\gallocv{1}, \ldots, \gallocv{\nagents})$ is a list of feasible bundles, one for each agent. Allocation $\allocs$ is \emph{feasible with respect to capacities} $\supplyvector$ if for all $\good \in \goodset$, we have that $\sum_{\agent \in \agentset} \alloc{\good}{\agent} \leq \supply{\good}$.

Each agent $\agent$ has a strict preference relation $\succ_\agent$ over the set $\choices_\agent$. 
We assume that $\ooo$ is the least preferred bundle for all agents. 
We denote the weak relation of $\succ_\agent$ by $\succeq_\agent$; i.e.,  $\x\succeq_\agent \y$ means  either $\x\succ_\agent \y$ or $\x=\y$.
Denote the preference profile of  all agents by $\succ:= (\succ_\agent)_{\agent \in \agentset}$.  The preference profile of a group of agents $\gsset$ is denoted by $\succ_\gsset := (\succ_\agent)_{\agent \in \gsset}$, and the preference profile of all agents outside $\gsset$ is denoted by $\succ_{-\gsset} := (\succ_\agent)_{\agent \notin \gsset}$. 
The \emph{type}  of agent $\agent$ is their preference relation $\succ_\agent$; we will sometimes denote the type of agent $\agent$ by $\ttype_{\agent}$ for notational clarity. Since the sets of feasible bundles and preference orderings are finite, the set of types is finite. Let $\typeset$ denote the set of agents' types (i.e., set of all strict preference orderings over bundles), and let $\numtypes = |\typeset|$ denote the number of agent types.

The \emph{economy} is a tuple $(N,\pi,M,\supplyvector{}, (\choices_\agent)_{\agent\in\agentset}, \succ)$. A (direct) \emph{mechanism} $\mechanism$ maps every preference profile  to a (random) allocation. 
Consider two economies $E$ and $E'$ that differ only in their permutations, denoted $\pi$ and $\pi'$ respectively, where $\pi$ and $\pi'$ are identical for agents $1, \ldots, k$. An \emph{online mechanism} is a mechanism that produces the same (random) allocation for agents $1, \ldots, k$ in both economies.

\section{\mech}

In a pseudomarket, agents are endowed with a \emph{budget} of tokens. They can spend these tokens to `buy' bundles at prevailing prices for the goods. Let  $\budgets = (\budget{1}, \ldots, \budget{\nagents})$ denote the vector of token budgets of the agents. In a \ceei, agents are allocated equal budgets of tokens \citep{varian1974equity}. However, in our setting, a \ceei\ might not exist.
We will therefore follow the spirit of \aceei\ introduced by \citet{budish2011combinatorial} and allow a slight perturbation of budgets as well as only approximate market-clearing in order to guarantee equilibrium existence.

\begin{definition}\label{def:aceeb} Fix an economy $(N,\pi,M,\supplyvector{}, (\choices_\agent)_{\agent\in\agentset}, \succ)$. The allocation $\optallocs=(\optgallocv{1}, \ldots, \optgallocv{\nagents})$, budgets $\optbudgets=(\optbudget{1},\ldots, \optbudget{\nagents})$ and prices $\optprices=(\optprice{1}, \ldots, \optprice{\ngoods})$ constitute an $\daisyeps-$\emph{dynamic approximate competitive equilibrium from equal incomes} (\mech)  if the following hold:
\begin{enumerate}[label=(\roman*)]
\item for all $\agentdef \in [\epsnn,\nagents]$, $\optgallocv{\agentdef} = \max_{\succ_\agentdef} \{\gallocv{\agentdef} \in \choices_\agentdef \text{ and } \optprices\cdot \gallocv{\agentdef} \leq \optbudget{\agentdef}\}$,\label{ACEEI1} 
\item  for all $\agentdef \in [\epsnn,\nagents]$, $1-\budgeteps \leq \optbudget{\agentdef} \leq 1$,\label{ACEEI2} 
\item for all $\agentdef \in [\epsnn,\nagents]$, $\good \in \goodset$,  $\sum_{\iter = 1}^{\agentdef} \optgalloc{\good}{\iter} \leq (1+\approxeps)\dfrac{\knum\supply{\good}}{\nagents}$, \label{ACEEI3}
\item for all $\agentdef \in [\epsnn,\nagents]$, $\good \in \goodset$,  if $\price{\good}>0$, then $\sum_{\iter = 1}^{\agentdef} \optgalloc{\good}{\iter}  \geq (1-\approxeps)\dfrac{\knum\supply{\good}}{\nagents} $.\label{ACEEI4}
\end{enumerate}
\end{definition}

\mech\ is parameterized by three error terms. The first error term, $\budgeteps$, determines the worst deviation from the equal budget. The second, $\sampleeps$, determines the fraction of agents for whom we do not impose any optimality, budget perturbation, or market-clearing conditions. Finally, $\approxeps$ controls the deviations from exact market-clearing. 
The first two parts in the definition of \mech\ are similar to \aceei: Part~\ref{ACEEI1} says that agents are allocated their most preferred bundle at the prevailing prices given their perturbed budgets; Part~\ref{ACEEI2} says that the agents' budget might be relaxed by at most $\budgeteps$.\footnote{This condition is slightly different from Definition 1.iii. in \citet{budish2011combinatorial} which bounds the increases in budgets. None of the results would be affected if we used his definition; our definition allows for a somewhat simpler exposition.} The final two parts of the definition extend \aceei~to the dynamic setting. Parts~\ref{ACEEI3} and~\ref{ACEEI4} say the markets for all objects clearly approximately \emph{in every period} except (possibly) for the first $\epsnn$ time periods. 
Note that 
$\frac{\agentdef\supply{\good}}{\nagents}$
denotes the proportional scaling of the capacity for each object when a $\frac{\agentdef}{\nagents}$--fraction of the agents has arrived. Therefore, the capacity of all goods is being used up at the rate of agent arrival which ensures that agents arriving earlier and later are treated fairly.

\subsection{Properties of \mech}
We first prove several results that highlight the fairness and efficiency properties of \mech.
While an allocation in  a \mech\ (or indeed in an \aceei)  might not be envy-free, we can show that whenever agent $i$ envies agent $i'$, it is possible to remove one object from $i'$'s bundle in such a way that $i$ no longer envies $i'$. 
\begin{definition} (\citealp{budish2011combinatorial}).
An allocation $\allocs$ is \emph{envy-free up to one object} (EF1) for a set of agents $\gsset$  if, for any $\agent, \agent' \in \gsset$, either  (i) $\gallocv{\agent}\succeq_\agent \gallocv{\agent'}$  or (ii) there exists some object $\good \in \goodsset$ such that $\gallocv{\agent}\succeq_\agent (\gallocv{\agent'}-\e^j)^+$.

\end{definition}

\begin{proposition}\label{prop:EF}
If $(\optallocs,\optbudgets,\optprices)$ is an $\daisyeps-$\mech\ with $\budgeteps< \frac{1}{m}$, then $\optallocs$ is envy-free  up to one object for the agents in $[\epsn \nagents,\nagents]$.
\end{proposition}
The proof of Proposition~\ref{prop:EF}  closely follows Theorem~3 in \citet{budish2011combinatorial} while adjusting for small differences in our equilibrium definitions.

 Our efficiency result is a version of the First Fundamental Theorem of Welfare which states that competitive equilibrium allocations are Pareto-efficient. Since capacities can be perturbed in our model, we first introduce a definition of Pareto efficiency that specifies the capacities of goods as well as the agents.

\begin{definition}
An allocation $\allocs$ that is feasible with respect to capacities $\supplyvector$ is (ex-post) \emph{Pareto-efficient for agents in $G$} if there is no other allocation which is feasible with respect to  $\supplyvector$ and is weakly preferred by all agents in $G$, with at least one strict preference.
\end{definition}

We consider the realized consumption of the goods once all the agents have arrived. We then perturb the capacities to match the realized consumption of the agents $\epsn \nagents,\ldots, \nagents$  and look at the efficiency of the allocations with respect to  these capacities.
\begin{proposition}\label{prop:PE}
Let $(\optallocs,\optbudgets,\optprices)$  be an $\daisyeps-$\mech\ of the economy and let $\optsupply{j}:=\sum_{\iter = \epsn \nagents}^{\nagents} \optgalloc{\good}{\iter}$  for all $\good \in \goodset$.
Then the allocation $\optallocs$ which is feasible with respect to capacities $\optsupplyvector$ is Pareto-efficient for agents in $[\epsn \nagents,\nagents]$.    
\end{proposition}

The proof of Proposition~\ref{prop:PE} is the standard revealed preference argument used for the First Welfare Theorem and is omitted.

\section{Implementation of \mech}
In this section, we introduce an \ocamlong\ (\ocam)    that produces a \mech\ with high probability when the market is sufficiently large and gives agents a strong incentive for truthtelling. To do so, we  define a new equilibrium  concept in which markets clear in expectation, which we call \emph{expected} \ceei\ (\eceei). In Section~\ref{sec:offline} we  formally define the \eceei\   and give a constructive proof for its existence. The construction uses the concept of a \emph{pseudoequilibrium} \citep{milgrom2009substitute} in an economy where agents' utilities depend on the quantity of tokens they have. As the tokens do not affect agents' preferences, a competitive equilibrium need not exist. The key idea in our construction is to approximate the ordinal preference ranking with cardinal utilities that are affected by money when the cost of the bundle is within $\epsilon$ of the budget.
The \ocam\ consists of two phases. In the first phase (which we call the \emph{sample} phase), we allocate bundles to arriving agents using a serial dictatorship. We then compute a set of prices and budgets for the sample agents that support a  competitive equilibrium in which markets clear in expectation. 
In the second phase, we use the prices and budgets from the first phase   to allocate bundles to  the remaining agents in an online fashion. We describe the mechanism in detail in Section~\ref{sec:online}; in Section~\ref{sec:prop} we describe the properties of the mechanism and state our main result. 

\subsection{\eceei\ }\label{sec:offline}
Consider a discrete random variable $\randombudget{}$ which we call a \emph{random budget}.
We say that a random budget $\randombudget{}$ is \emph{$\epsilon$-close-to-1} if all realizations of $\randombudget{}$ are between $1-\epsilon$ and $1$. 
For any agent $\agent$,  price vector $\prices$ and random budget $\randombudget{\agent}$, define 
\begin{equation}\label{eq:xhat}
\optbundle{\agent}(\prices,\randombudget{\agent}) := \left\{\max_{\succ_{\agent}} \{\gallocv{} \in \choices_{\agent} \text{ and } \prices\cdot \gallocv{} \leq \budget{\agent} \} \text{ with probability } \budget{\agent} \sim \randombudget{\agent}\right\}.
\end{equation}
 We call $\optbundle{\agent}(\prices,\randombudget{\agent})$ a \emph{random optimal bundle} for agent $\agent$. The realizations of $\optbundle{\agent}$ are the   optimal bundles for agent $\agent$ at prices $\prices$ when $\agent$'s budget is drawn from the distribution $\randombudget{\agent}$. 
Denote agent $i$'s \emph{expected optimal bundle} by $\expect{\optbundle{\agent}(\prices,\randombudget{\agent})}$, where the expectation is over $\randombudget{\agent}$.  
Using these definitions, we state our equilibrium concept.

\begin{definition}Fix an economy $(N,\pi,M,\supplyvector{}, (\choices_\agent)_{\agent\in\agentset}, \succ)$. Given an $\epsilon>0$,  random $\epsilon$-close-to-$1$ budgets $(\optrandombudget{1},\ldots,\optrandombudget{\nagents})$,  prices $\optprices = (\optprice{1},\ldots,\optprice{\ngoods})$ and the random  allocation $\optbundlev = (\optbundle{1},\ldots,\optbundle{\nagents})$ comprise an \emph{$\epsilon$-expected competitive equilibrium from equal incomes} ($\epsilon$-\eceei) 
if for every good $\good$, 
\begin{enumerate}[label=(\alph*)]
\item $\sum_{\agent \in \agentset} \expect{\optbundle{\agent}(\optprices,\optrandombudget{\agent})}_\good \le \supply{\good}$,
\item if $\price{\good}>0$, then  $\sum_{\agent\in \agentset} \expect{\optbundle{\agent}(\optprices,\optrandombudget{\agent})}_\good= \supply{\good}$.
 \end{enumerate}  
\end{definition}

The definition of the \eceei\ is intuitive: if each agent consumes their expected optimal bundle given their random budgets then the markets for goods clear exactly. Before proving the existence of 
an $\epsilon$-\eceei\ for any $\epsilon >0$, we require some more definitions. 

For each agent $\agent$ and a feasible bundle $\bundle$, let  $r_\agent(\bundle)$ be some \emph{numerical utility}  that is consistent with the preference order $\succ_\agent$. 
We also set  $r_\agent(\emptybundle)=0$ and $r_\agent(\bundle)=-\infty$  for any infeasible bundle $\bundle$.
For any  $\epsilon>0$, define the following  \emph{auxiliary utility} function for each agent $\agent$.\footnote{Thus,
the auxiliary utility of a feasible bundle 
is  $-\infty$ if its cost is at least 1, while for  an infeasible bundle, the utility is always $-\infty$ regardless of price.}

\begin{equation}\label{eq:auxiliary}
  \util^{\epsilon}_\agent(\bundle,\prices)=\begin{cases}
    r_\agent(\bundle)+\min\{0,\log\frac{1-\prices\cdot \bundle}{\epsilon}\} , & \text{if $\prices\cdot \bundle<1$}.\\
    -\infty , & \text{otherwise}.
  \end{cases}
\end{equation}




\begin{figure}[tbp]
    \centering
    \includegraphics[width=3.5in]{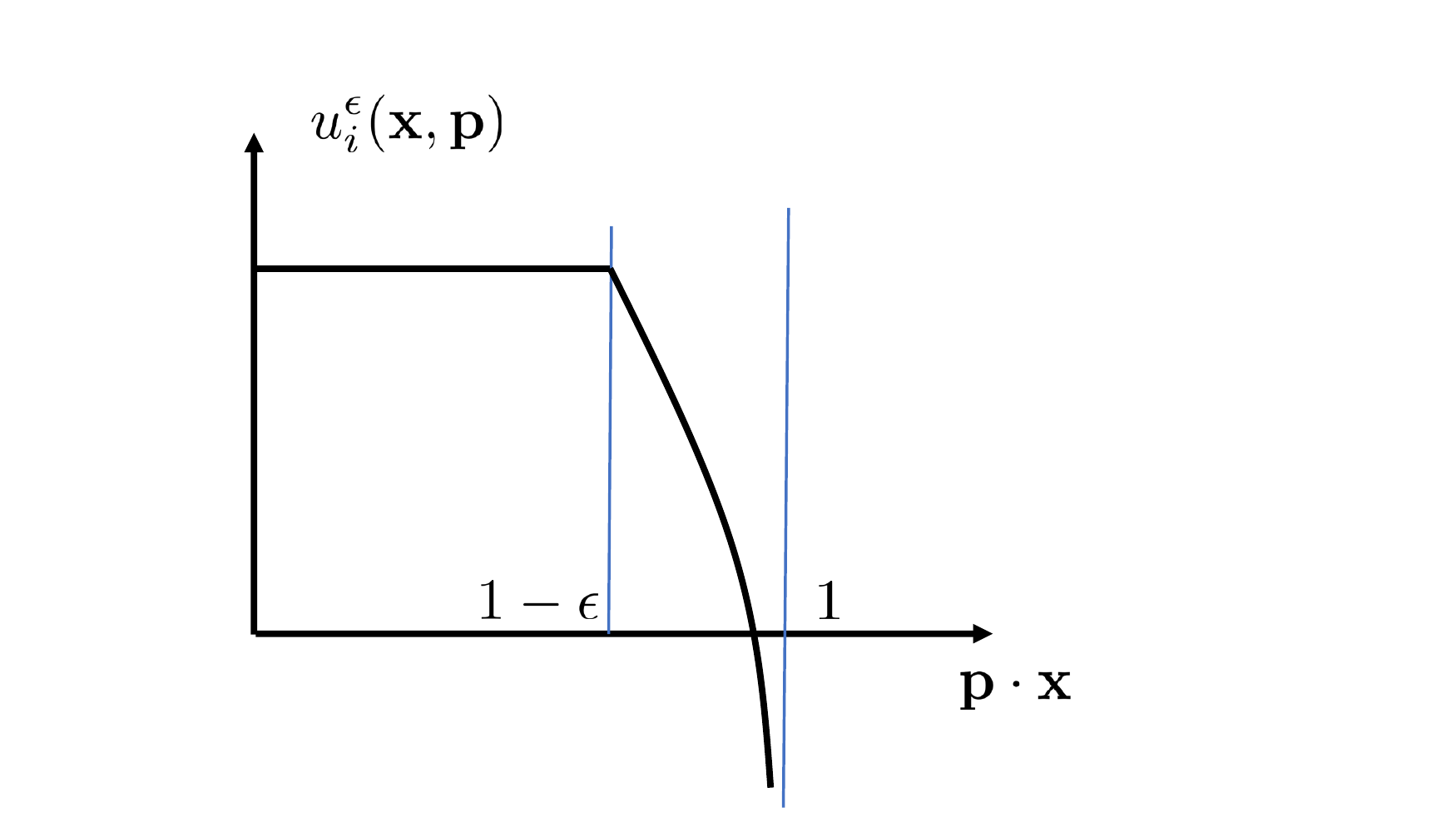}
    \caption{Auxiliary utility  \label{fig:aux}}
      
\end{figure}
The auxiliary utility is illustrated in Figure~\ref{fig:aux}.
Note that the auxiliary utility only differs from 
$r_{\agent}(\bundle)$ when the cost of the bundle $\prices \cdot \bundle$ is at least $1-\epsilon$. 
We now define a pseudoequilibrium  for agents with auxiliary utilities. In an auxiliary economy, ordinal preferences of agents are simply replaced by auxiliary utilities $\util^{\epsilon}_{\agent}(\bundle,\prices)$, and budgets are no longer relevant. For a price vector $\prices$, define $Ch_{\agent}(\prices)=\arg\,\max_\bundle\{\util^{\epsilon}_{\agent}(\bundle,\prices)\}$ and denote by $conv(A)$ the convex hull of set $A$.

\begin{definition} 
Fix an auxiliary economy. The allocation $\optallocs=(\optgallocv{1}, \ldots, \optgallocv{\nagents})$ and prices $\optprices=(\optprice{1}, \ldots, \optprice{\ngoods})$ constitute a \emph{pseudoequilibrium}
      if $\optgallocv{\agent}\in conv(Ch_{\agent}(\prices))$ for all $\agent \in \agentset$
and $\sum_{\agent \in \agentset}\optgallocv{\agent} \leq \supplyvector $ with equality for every good $\good \in \goodset$ with $\price{\good} \neq 0$.
\end{definition}
In words, a pseudoequilibrium is a relaxation of competitive equilibrium in settings with indivisible goods where all agents' preferences are convexified by replacing the choice correspondence $Ch_{\agent}(\prices)$ of each agent $\agent$ with its convex hull $conv(Ch_{\agent}(\prices))$. We are now ready to state our main result for this section.

\begin{theorem}\label{theo:ECEEI}
For every  $\epsilon>0$, there exists a pseudoequilibrium  with auxiliary utilities $\{\util^{\epsilon}_{\agent}(\bundle,\prices)\}_{\agent \in \agentset}$.
Furthermore, suppose that allocations $\optallocs$ and prices $\optprices$ form such a pseudoequilibrium. Then one can construct  $\epsilon$-close-to-1 random budgets 
$(\optrandombudget{1},\ldots,\optrandombudget{\nagents})$, such that these budgets, random allocations $\{\optbundle{\agent}(\optprices,\optrandombudget{\agent})\}_{\agent  \in \agentset}$
and prices $\optprices$ form an $\epsilon$-\eceei. Moreover, in the pseudoequilibrium agents of the same type have the same allocation and, as a result, in the constructed $\epsilon$-\eceei, agents of the same type have the same random budget.
\end{theorem}



\subsection{The \ocamlong\  (\OCAM)}\label{sec:online}
In this section, we describe our mechanism; the pseudocode appears in the online appendix. The initial input to the \ocam\ is a set $\goodsset$ of goods  with capacity $\supplyvector{}$ 
and three error terms, $\epsilons$ (as in Definition~\ref{def:aceeb}). 
We  set the sample size  to be $\realsample$, where $\epss = \frac{\epsf \epsn}{4}$. Recall that $\epsf$  determines the deviation from exact market-clearing and $\epsn$ is the fraction of agents for whom we do not impose any conditions. In this way, the \ocam\ enables a parameterized tradeoff between these two guarantees. The
\ocam\ is a direct online mechanism: it elicits agents' types and allocates each agent a bundle immediately upon arrival. 
When a sample agent arrives, the \ocam\ allocates the agent a bundle using serial dictatorship with the capacities scaled down proportionally to the initial sample size. That is, the \ocam\ allocates the agent their favorite bundle such that for any good $\good$, the total amount allocated of good $\good$ is at most $\epss \supply{\good}$. The \ocam\ also allocates the agent an arbitrary budget from $[1-\epsb, 1]$ (this budget is never used, but it is technically required for the definition of \mech). 

Once all of the sample agents have arrived, we compute an  $\epsb$-\eceei~ for the sample. An $\epsb$-\eceei\ is defined by random allocations, budgets, and prices, but we only require the  budgets $\randombudget{1}, \ldots, \randombudget{\realsample}$ and prices $\prices$. Note that these are \emph{not} the budgets that the \ocam\ actually assigned to the sample agents; they  will only be used for future assignments. We now define a function that maps agent types to random budgets. For any agent type $\ttype$, if there was an agent $\agent \in \{1, \ldots, \realsample\}$ of type $\ttype$, set $\budgetfunc(\ttype) = \randombudget{\agent}$. In other words, when an agent arrives, if their (declared) type was in the sample, set their budget to be identical to the (perfunctory) budget of an agent of the same type. 
Otherwise, set their budget arbitrarily between  $1-\budgeteps$ and $1$.
Define the allocation rule for the \mech\ as follows: for agent $\agent  = \realsample+1, \ldots, \nagents$, draw $\budget{\agent} \sim \budgetfunc(\ttype_\agent)$, where $\ttype_{\agent}$ is agent $\agent$'s type and set $\allocv{\agent} = \max_{\succ_{\agent}}\{\bundle{}: \bundle{} \in \choices_{\agent} \text{ and } \prices\cdot \bundle{} \leq \budget{\agent}\}$ (i.e., allocate the agent their favorite bundle given the drawn budget).
Using this allocation rule, allocate bundles and budgets to agents $\realsample+1, \ldots, \nagents$ upon arrival. 

\ignore{
 
\begin{theorem}\label{thm1}
For any $\epsf, \sampleeps,\budgeteps,\ngoods, \nagents, \numtypes>0$, if Assumption~\ref{ass1} holds, there exists an algorithm that outputs an allocation $(\allocv{1}, \ldots, \allocv{\nagents})$, budgets $(\budget{1},\ldots, \budget{\nagents})$ and prices $(\price{1}, \ldots, \price{\ngoods})$ that constitute a dynamically  $(\sampleeps,\budgeteps,\epsf)-$approximate competitive equilibrium with equal budget with probability at least $1- \frac{1}{\nagents}$.
\end{theorem}

}

\subsection{Properties of the \ocam}\label{sec:prop}

In this section, we demonstrate three desirable properties of the \ocam: efficiency, fairness, and strategyproofness. The first two properties stem from the fact that the \ocam\ implements \mech\ with high probability (Theorem~\ref{thm1}).

There is a variety of existing definitions of strategyproofness for random mechanisms that use only ordinal preference information. One such definition demands that truthtelling be a dominant strategy for every realized outcome of the mechanism (we refer to it as `ex post' strategyproofness). In our setting, the only efficient mechanisms that satisfy ex post strategyproofness are serial dictatorships \citep{klaus2002strategy}. Weaker (or `interim') notions of strategyproofness only require non-manipulability whenever agents' preferences over lotteries are represented by von Neumann-Morgenstern utility functions or by the stochastic dominance partial order \citep{bogomolnaia2001new}.\footnote{To see that ex post strategyproofness is a stronger non-manipulability condition, note that it requires that truthtelling be a dominant strategy for \emph{any} utility function representation of preferences over lotteries while interim strategyproofness notions restricts which functions can represent preferences.}


Here, we use a notion of ex post strategyproofness and require that an agent cannot significantly improve \emph{any} realization of the random allocation by misrepresenting their type. We avoid the incompatibility of ex post strategyproofness with Pareto efficiency and non-dictatorship by only requiring that manipulations be sufficiently profitable and efficiency be approximate. Furthermore, 
our mechanism disincentivizes any {group}  of  agents from misrepresenting their type.


We will need some additional notation. Given an economy and a random mechanism $\mechanism$, let $\mechanism_i(\succ)$ denote the mapping of the agents' preference profile to agent $i$'s random allocation, and let $R(\mechanism_i(\succ))$ denote the set of all bundles with positive probability in the random allocation $\mechanism_i(\succ)$. Using this notation, we can now define
 group-strategyproofness up to one object.

\begin{definition}\label{defn:gsp}
A mechanism $\mechanism$ is (weakly) \emph{group-strategyproof up to one object}   if for any group of agents $\gsset$ and any two preference profiles of these agents $\succ_\gsset$ and  $\succ'_\gsset$ at least one of the following conditions is true.
\begin{enumerate}[label=(\roman*)]
    \item there exists an agent $\agent \in G$
such that
for any $\gallocv{\agent}\in R(\mechanism_i(\succ_\gsset ,\succ_{-\gsset}))$  and  $\gallocv{\agent}'\in R(\mechanism_i(\succ'_\gsset ,\succ_{-\gsset}))$, it holds that $\gallocv{\agent} \succeq_\agent \gallocv{\agent}'$,
\item for all $\agent \in G$,  $\gallocv{\agent}\in R(\mechanism_i(\succ_\gsset ,\succ_{-\gsset}))$, and   $\gallocv{\agent}'\in R(\mechanism_i(\succ'_\gsset ,\succ_{-\gsset}))$, then there exists  $\good \in \goodsset$ such that  $\gallocv{\agent}\succeq_\agent (\gallocv{\agent}'-\mathbf{e}^\good)^+$.
\end{enumerate}
\end{definition}

Definition~\ref{defn:gsp} says that if a group of agents $G$ misreports their types, then either there exists an agent  $\agent \in G$ that is weakly worse off, or for \emph{every agent} $\agent \in G$, any realized allocation that agent  $i$ obtains by misreporting their  type can be made weakly worse than any realized allocation under truthtelling by removing some object.





%

Next, we introduce two assumptions---on the arrival order and on market size---that allow us to show the existence of a mechanism that implements a DACEEI.
First, we require that the agents arrive at random.
\begin{assumption}\label{ass0}
    Agents arrive according to  permutation selected uniformly at random  from the set of  all permutations on $[n]$.
\end{assumption}
Note that we make no assumption on the distribution of types; our results hold for  `arbitrarily bad'  distributions.  
Second, we require that the capacities of the goods be sufficiently large.
\begin{assumption}\label{ass1}
$\min_{\good} \{\supply{\good}\} \geq  \dfrac{70\left(\numtypes\log{(\realsample)} + \log{(\ngoods)}+\sqrt{\nagents}\log{\nagents}\right)}{\epss \epsf^2}$, where $\epss = \frac{\epsn\epsf}{4}$.
\end{assumption}

We do not attempt to optimize the constant in Assumption~\ref{ass1}, preferring clarity in the proofs to a better constant. The $\sqrt{\nagents}$ term in Assumption~\ref{ass1} is only required to ensure that the allocation is approximately fair and efficient along the entire sequence of agents' arrival. If one is willing to relax this assumption and only require that these desiderata hold for the final allocation, we can discard the $\sqrt{\nagents}$ term (alternatively, if the bound on $\supplyvector$ holds only for the other terms in the numerator, the result still holds for the final allocation). For comparison, the capacity assumption of~\citet{AgrawalOnline} 
is $\min_{\good} \{\supply{\good}\} \geq  \Omega\left(\frac{\ngoods \log{(\nagents/\epsilon)}}{\epsilon^3}\right)$.
When  $\numtypes$ is sufficiently small, as is the case for all of the applications described in the introduction, our bound is only worse than theirs by a factor of $\epsilon$. We also note that the assumption is independent of $\epsb$, hence we can make $\epsb$ arbitrarily small. 
 Our main result on  the properties of the \ocam\ is the following.
\begin{theorem}\label{thm1}
{For any economy $(N,\pi,M,\supplyvector{}, (\choices_\agent)_{\agent\in\agentset}, \succ)$,  $\epsf, \sampleeps>0$, and $0<\budgeteps< \frac{1}{m}$,} the \ocam\,~is: 
   \begin{enumerate}[label=(\alph*)]
   \item group-strategyproof up to one object,\label{thm1a}
   \item envy-free up to one object for a $(1-\epss)$ fraction of the agents, where $\epss = \frac{\epsn\epsf}{4}$. \label{thm1b}
   \end{enumerate}
Furthermore,   when   Assumptions~\ref{ass0} and \ref{ass1} hold, the \ocam\ outputs a tuple $(\optallocs,\optbudgets,\optprices)$   
that constitutes an $\daisyeps-$\mech\ with probability at least $1- \frac{1}{\nagents}$. 
\end{theorem}

Note that neither the fairness nor the strategic properties depend on the order of arrival or on market size. Indeed, Assumptions~\ref{ass0} and~\ref{ass1} are only needed to prove that approximate market-clearing  holds in almost every period as required by the \mech.

\section{Conclusion}
The \ocam\  is a promising mechanism for complex, dynamic allocation problems without money. It finds an attractive compromise among fairness, efficiency and incentives for truthtelling and uses up the capacity of each good at a constant rate.

There are several possible fruitful directions for further work. One direction is theoretical: one might wish to relax the assumption that the arrival order is random (Assumption~\ref{ass0}); however, this would require a significant adaptation of our techniques. 
Theorem~\ref{thm1} suggests that if the initial sample represents the population well, then equilibrium prices computed from initial sample will be  a good guide to an efficient allocation for the remaining agents. In richer environments with a non-stationary arrival order, one could, for example, consider using adaptive sampling to ensure that the samples are representative.
Another direction would be to test the \ocam\ empirically or 
even implement it in practice. 

\bibliographystyle{chicago}
\bibliography{ref}

\newpage

\appendix
\begin{center}
{\bf \huge APPENDIX}
\end{center}
\vspace{0.4cm}

\section{Proof of Theorem~\ref{theo:ECEEI}: Existence of \eceei}\label{app:eceei}

\cite{nguyen2021delta} establish the following result on the  existence of a pseudoequilibrium in a more general setting.

 \begin{lemma}[\citealp{nguyen2021delta}]\label{lemma:pseudo}
Let $\choices_{\agent}$ denote the finite set of bundles that agent $j\in N$ can feasibly consume, $\emptybundle \in \choices_{\agent}$. Each agent $\agent$'s utility function $\util_{\agent}(\bundle,\prices)$ satisfies
\begin{itemize}
    \item $\util_{\agent}(\emptybundle,\prices)=0$,
    \item $\util_{\agent}(\bundle,\prices)=-\infty$ for $\bundle\notin \choices_{\agent}$
    \item  $\util_{\agent}(\bundle,\prices)$ is continuous  in $\prices \in \mathbb{R}^m$ for each  $\bundle\in \choices_{\agent}$ and
    \item there exists some constant  $C>0$ such that if $\prices\cdot\bundle \ge C$, then, $\util_{\agent}(\bundle,\prices)<0$.
\end{itemize}
Then, there exists a pseudoequilibrium.
\end{lemma}

We use this result to prove Theorem~\ref{theo:ECEEI}.
\proof{Proof of Theorem~\ref{theo:ECEEI}. }
The utilities $\util^{\epsilon}_{\agent}(.,.)$ in the theorem statement  satisfy Lemma~\ref{lemma:pseudo}, therefore a pseudoequilibrium exists.  We use this pseudoequilibrium to construct an $\epsilon$-\eceei, as follows.


Let  $\optprices$ and $\{\optgallocv{\agent}\in conv(Ch_{\agent}(\optprices))\}$ be a pseudoequilibrium under the utilities $\util^{\epsilon}_{\agent}(.,.)$. For  $\y\in Ch_{\agent}(\optprices)$, define a budget $b_\y$ as follows:

\begin{equation}\label{eq:budget}
  b_\y=\begin{cases}
    1-\epsilon  & \text{if $\optprices\cdot \y\le 1-\epsilon$}.\\
    \optprices\cdot \y  & \text{otherwise}.
  \end{cases}
\end{equation}
Notice  that $\optprices\cdot \y<1$ because  $\y\in Ch_{\agent}(\optprices)$, $\util^{\epsilon}_{\agent}(\y,\optprices)\ge 0$. Thus, for every $\y\in Ch_{\agent}(\optprices)$, $1-\epsilon \le b_\y<1$.
Next we show that $\y\in Ch_{\agent}(\optprices)$ implies that $\y$ is the best bundle among all the bundles with a cost at most $b_\y$. That is:
\begin{equation}\label{eq:optimal}
\y=\max_{\succ_{\agent}}\{\gallocv{} \in \choices_{\agent} \text{ and } \optprices\cdot \gallocv{} \leq b_\y\}.
\end{equation}
This is true because in the case that $b_\y= 1-\epsilon$,  the auxiliary utility of bundles with a cost at most $b_\y$ is equal to the  numerical value $r_{\agent}(.)$ associated with its ordinal ranking. $\y\in Ch_{\agent}(\optprices)$ implies that $r_{\agent}(\y)$ is the highest among these bundles.
Now, if $b_\y=\optprices\cdot \y >1-\epsilon$, then  for any bundle $\bundle$ such that  $r_{\agent}(\bundle)>r_{\agent}(\y)$ we have $\optprices\cdot \bundle>\optprices\cdot \y$, otherwise the  
auxiliary utility of bundle $\bundle$ is strictly higher than that of  $\y$,
contradicting the fact that $\y\in Ch_{\agent}(\optprices)$.

Since $\optgallocv{\agent}\in conv(Ch_{\agent}(\optprices))$,  $\optgallocv{\agent}$ can be expressed as the  expectation  of a lottery $\mathcal Z$ over the bundles in $Ch_{\agent}(\optprices)$.  Let  
$\optrandombudget{\agent}$ be the corresponding random budget defined as in \eqref{eq:budget} for $\y$ drawn from $\mathcal Z$.

In this construction, every realization of $\optrandombudget{\agent}$ is $b_\y$ for $\y\in Ch_{\agent}(\optprices)$, thus it is between $1-\epsilon$ and 1.
Moreover, because of \eqref{eq:optimal}, the random optimal bundle with respect to budget $\optrandombudget{\agent}$ is the lottery $\mathcal Z$ over $Ch_{\agent}(\optprices)$ with the average equal to $\optgallocv{\agent}$. The market-clearing condition of the pseudoequilibrium implies the market clear condition of the \eceei.
Finally, agents of the same type have the same auxiliary utility, and we can select a pseudoequilibrium such that they have the same fractional allocation, which implies that agents of the same type have the  same random budgets in the $\epsilon$-\eceei.  
\proofen

\section{\eceei\ implies  \aceei}

In this section, we show that 
 the existence of \eceei~implies the existence of an approximate competitive equilibrium from equal incomes with the same bound on the {excess} demand as in \cite{budish2011combinatorial}. In particular we show the following result.

 \begin{proposition}~\label{prop:budish}
     There is a realisation of  \eceei\ which has the excess demand bounded in $\ell_2$--norm at most  $\sqrt{\sigma m/2}$, where $\sigma$ is the size of the maximum bundle consumed by any agent.
 \end{proposition}

To show this, we use the following improved bound for the Shapley--Folkman theorem \citep{budish2020improved} in order to obtain an  \aceei. 
\begin{lemma}[Theorem 3.1 in \citealp{budish2020improved}] \label{theo:bush} 
If $S_1,..,S_n$ are compact subsets of $\mathbf{R}^m$, if ${\bf c}\in conv(S_1+..+S_n)$, and $D$ is the maximum diameter of $S_i$, then there exists $\x_i\in S_i$ such that 
    $$
    ||{\bf c}-\sum_i \x_i||_{\ell_2} \le D\sqrt{m}/2.
    $$
\end{lemma}

\proof{Proof of Proposition~\ref{prop:budish}. }
Let $\prices, \randombudget{}$ be an $\epsilon$-\eceei.  Apply Lemma~\ref{theo:bush}, where $S_i=\optbundle{\agent}(\prices,\randombudget{\agent})$ and ${\bf c}$ is the capacity vector.  If all agents' feasible bundles are of size at most $\sigma$, then the diameter of $S_i$ is at most $\sqrt{2\sigma}$. We obtain the existence of 
$\x_i\in S_i$ such that $ ||{\bf c}-\sum_i \x_i||_2 \le \sqrt{\sigma m/2}.$ For each $\x_i\in \optbundle{\agent}(\prices,\randombudget{\agent}) $, there is a realization $b_i$ of $\randombudget{\agent}$ such that 
$\x_i$ is the optimal choice of agent $i$ under budget $b_i$. Thus the allocation $\{\x_1,..,\x_n\}$ corresponds to an approximate equilibrium, where the budget of each agent is perturbed by at most $\epsilon$.
\proofen

\section{Concentration Inequalities}
We use the following concentration inequalities.

\begin{theorem} [Chernoff Bound]\label{Chernoff}
Let $X_1, X_2, \ldots, X_n$ be independent random variables with  $X_i = 1$ with probability $p_i$ and $X_i = 0$ with probability $1-p_i$. Define $X = \sum_{i=1}^{N} X_i$, and  $\mu = \expect{X} = \sum_{i=1}^{N} p_i$. Then for $\epschernoff   \in (0,1)$,
\begin{enumerate}
\item $\Pr[X-\mu \geq \epschernoff \mu ] \leq \exp\left(-\mu\epschernoff ^2/3\right)$,
\item $\Pr[\mu-X \geq \delta \mu ] \leq \exp\left(-\mu\epschernoff ^2/2\right)$.
\end{enumerate}
\end{theorem}

The following is adapted from~\cite{bardenet2015concentration}. They only prove a single-sided bound (i.e., only the first inequality), but it is not too difficult to see that the proof holds for the complementary inequality.\footnote{Their proof hedges on showing that $Z^*_k = \frac{1}{N-k}\sum_{t=1}^{k}(X_t-\mu)$ is a martingale: $\expect{Z^*_k|Z^*_{k-1}, \ldots,Z^*_1} = Z^*_{k-1}$. It is straightforward to adapt their proof to show that $Z_k = \frac{1}{N-k}\sum_{t=1}^{k}(\mu-X_t)$ is also a martingale, and the result follows from this. } A similar (two-sided) bound can be derived from~\cite{Serfling}. We note that the bound of~\cite{bardenet2015concentration} is tighter than the one below, but for our values of $\nagents$  and $\nsample$ (which represent the number of agents and sample size respectively), the improvement is negligible.

\begin{theorem} [Hoeffding-Serfling Inequality] \label{Hoeffding}
 Let  $\hoefset = \{\hoef_1, \ldots, \hoef_\nagents\}$ be a finite set of $\nagents$ elements, where for $\agent \in \agentset$, $\hoef_\agent \in [0,1]$.  Let $\hoefrand_1, \ldots, \hoefrand_\nsample$ be a random sample drawn without replacement from $\hoefset$. Define $\avgpref = \frac{1}{\nagents}\sum_{\agent = 1}^{\nagents} \hoef_\agent$.  Then for $\epsHoef>0$, 
 \begin{enumerate}
 \item $\Pr\left[\max_{ \nsample \leq \kn \leq \nagents} \sum_{\agent = 1}^{\kn} \hoefrand_\agent - \kn \avgpref > \kn \epsHoef  \right] \leq \exp{\left(-2\nsample \epsHoef^2\right)},$
\item $\Pr\left[\max_{ \nsample \leq \kn \leq \nagents} \sum_{\agent = 1}^{\kn} \hoefrand_\agent - \kn \avgpref < \kn \epsHoef  \right] \leq \exp{\left(-2\nsample \epsHoef^2\right)}.$
\end{enumerate}
\end{theorem}
The following bound is tighter than the Hoeffding bound when the variance of $\hoefset$ is bounded. We use a version that appears in~\citep[e.g.,][]{aw}.

\begin{theorem}[Hoeffding-Bernstein Inequality]\label{thm:Bern} Let  $\hoefset = \{\hoef_1, \ldots, \hoef_\nagents\}$ be a finite set of $\nagents$ elements, where for $\agent \in \agentset$, $\hoef_\agent \in [0,1]$.  Let $\hoefrand_1, \ldots, \hoefrand_\nsample$ be a random sample drawn without replacement from $\hoefset$. Define $\avgpref = \frac{1}{\nagents}\sum_{\agent = 1}^{\nagents} \hoef_\agent$.  Then  for $\epsBern>0$,
	$$\Pr\left[ \left|\sum_{\agent=1}^{\nsample} \hoefrand_\agent -  \nsample \avgpref \right| \geq \epsBern\right]\leq 2\exp\left( \frac{-\epsBern^2}{2\nsample \sigma^2_\nagents+\epsBern} \right),  $$
	where $\sigma^2_\nagents = \frac{1}{\nagents}\sum_{\agent=1}^\nagents (\hoef_\agent-\avgpref)^2$.
\end{theorem}

\section{Proof of Theorem~\ref{thm1}} \label{app:proofthm}

We first prove Parts~\ref{thm1a} and~\ref{thm1b} of Theorem~\ref{thm1}, rephrased as the following lemma.
\begin{lemma}\label{lemmathm1}
For any economy $(N,\pi,M,\supplyvector{}, (\choices_\agent)_{\agent\in\agentset}, \succ)$,  $\epsf, \sampleeps>0$ and $0<\budgeteps< \frac{1}{m}$, 
 the \ocam\ is: 
   \begin{enumerate}[label=(\alph*)]
   \item group-strategyproof up to one object,\label{asdasd1a}
   \item envy-free up to one object for a $(1-\epss)$ fraction of the agents, where $\epss = \frac{\epsn\epsf}{4}$. \label{asasd1b}
   \end{enumerate}
\end{lemma}

\proof{Proof of Lemma~\ref{lemmathm1}.} For the first $\realsample$ agents (the sample), the mechanism is a serial dictatorship and therefore group-strategyproof. It remains to show that when $\epsb<\frac{1}{m}$, the allocations of the agents arriving after the sample are  envy-free, and group-strategyproof up to one object.

First observe that, in the OCAM, the allocations of agents arriving after the sample  phase
do not depend on the reported type of others. For these agents, 
when misreporting their type from $t$ to $t'$, they will receive an allocation that  agents of type $t'$ receive.  
Thus, ex-post envy-freeness up to one object implies group-strategyproofness up to one object.

It remains to show that the allocation is ex-post envy-free up to one object for the agents arriving after the  sample phase. Let $i,i'$ be two such agents, and let $(\x_i, b_i)$ and   $(\x_{i'}, b_{i'})$ be the ex-post allocation and budget of agent $i$ and $i'$, respectively.

Assume the contrary: that taking any object out of $\x_{i'}$, agent $i$ still prefers it to their current bundle $\x_i$, then it must be that the cost of that bundle ($\x_{i'}$ with the object removed) is higher than $b_i$. Thus,  we have
$$b_i< \prices \cdot (\x_{i'}-\e^j) \text{ for every  good $j$ contained in bundle $\x_{i'}$}. $$
Summing up these inequalities for all such  $j$, we obtain  
$$
\sigma\cdot  b_i< (\sigma -1)\cdot  \prices \cdot \x_{i'}, \text{ where $\sigma$ is the size of bundle } \x_{i'}.
$$
On the other hand, because $\x_{i'}$ is the bundle consumed by agent $i'$, we have $\prices \cdot \x_{i'}\le b_{i'}$. Thus, $\sigma\cdot  b_i< (\sigma -1)\cdot b_{i'}$, which implies
$$
\frac{b_i}{b_{i'}}<\frac{\sigma-1}{\sigma}\le \frac{m-1}{m}.
$$
This is a contradiction because both $b_i,b_{i'} \in (1-\frac{1}{m},1)$.
\proofen

 To complete the proof of Theorem~\ref{thm1}, we need to show that   when   Assumptions~\ref{ass0} and \ref{ass1} hold, the mechanism outputs a tuple $(\optallocs,\optbudgets,\optprices)$   
that constitutes an $\daisyeps-$\mech\ with probability at least $1- \frac{1}{\nagents}$. 
We first require some additional notation. 

For any subset of size $\realsample$ of the agents, we define a triple $\triple = (\outputtypeset, \budgetfunc, \prices)$ where $\outputtypeset$ is an unordered set (with repetition) of the types of the agents in the subset, $\budgetfunc$ is a function mapping agent types  to 
 a random budget, and $\prices$ is a price vector for $\ngoods$ objects ($\budgetfunc$ and $\outputtypeset$ are as defined in the pseudocode of Mechanism~\ref{alg:one}.) The triple $\triple$ will be used to define allocations for all of the agents. However, the allocations defined by $\triple$ and those generated by the mechanism are not (necessarily) the same:   the allocations will be identical  for agents $\realsample+1, \ldots, \nagents$,  but  for agents $1, \ldots, \realsample$, they  may not be. For any agent $\agent$ and triple $\triple = (\outputtypeset, \budgetfunc, \prices)$, let  $\ballocv{\agent}(\triple)\in [0,1]^{2^\ngoods}$ denote the randomized allocation of agent $\agent$ that is generated by $\triple$, where $\balloc{\agent}{\bundle}$ denotes the probability that agent $\agent$ is allocated bundle $\bundle$. 
  The value of $\ballocv{\agent}$ is  uniquely determined by $(\outputtypeset, \budgetfunc, \prices)$.  Let $\rallocv{\agent}(\triple)\in [0,1]^{\ngoods}$ be  such that $\ralloc{\good}{\agent}(\triple)$ denotes the  probability that agent $\agent$ is allocated object $\good$ (that is, $\ralloc{\good}{\agent}(\triple) = \sum_{\bundle: \good \in \bundle}\balloc{\agent}{\bundle}(\triple)$); let $\rvrallocv{\agent}$ be the corresponding random variable (with the randomness over the choice of $\triple$).
Finally, let $\rvallocv{\agent} (\triple)$ be the random variable denoting the realization of $\ballocv{\agent}(\triple)$; that is, $\rvalloc{\good}{\agent}(\triple) \in \{0,1\}$, where $\rvalloc{\good}{\agent}(\triple) = 1$ with probability $\ralloc{\good}{\agent}(\triple)$. 
Note that the events $\rvalloc{\good}{\agent}(\triple) = 1 $ and $\rvalloc{\agent}{\good'}(\triple)= 1$ are dependent,  but for any $\agent \neq \agent'$, $\good, \good'$, $\rvalloc{\good}{\agent}(\triple)= 1 $ and $\rvalloc{\agent'}{\good'}(\triple)= 1 $ are independent.  Overloading the notation, let $\rvallocv{\agent}$ denote the random variable for the allocation of agent $\agent$, where the sample space includes randomness from both the random arrival order and the budget selection.
In Sections~\ref{d1} and~\ref{d2}, we bound the complete randomized and deterministic allocations defined by the triple $\triple$ (where by `complete' we mean with respect to all of the agents' allocations). In Sections~\ref{d3} and~\ref{d4} we use these results to prove bounds on the allocations generated by~\ocam.

\subsection{Step 1: Bounding the  randomized allocation generated by the triple \texorpdfstring{$\triple$}{q}. }\label{d1}

For the first step, we fix an arbitrary triple $\triple = (\outputtypeset, \budgetfunc, \prices)$ and define a sample to be bad with respect to this triple as follows:

\begin{definition}
We say that a sample $\sample$ is \emph{bad} for triple $\triple = (\outputtypeset, \budgetfunc, \prices)$ and object $\good$ if  either 
\begin{enumerate}[label=(\roman*)]
\item $\sum_{\agent \in \sample}\ralloc{\good}{\agent}(\triple) \leq \epss \supply{\good}$ and $\sum_{\agent \in \agentset}\ralloc{\good}{\agent}(\triple) >  \left(1+\epsfa\right) \supply{\good}$ or
\item $\sum_{\agent \in \sample}\ralloc{\good}{\agent}(\triple) \geq  \epss \supply{\good}$ and $\sum_{\agent \in \agentset}\ralloc{\good}{\agent}(\triple) <  \left(1-\epsfa\right)\supply{\good}$.
 \end{enumerate}
\end{definition}

\begin{lemma}\label{lemma:rnd}
For any $\epsilons$, if Assumption~\ref{ass1} holds, then the randomized allocation $\rvrallocv{}$ and prices $\prices$ generated  by the triple $\triple$ satisfy the following: 
\begin{enumerate}
\item The probability that there exists some object $\good \in \goodset$ such that $\sum_{\agent \in \agentset}\rvralloc{\good}{\agent} >\left(1+\epsfa\right) \supply{\good}$ is at most $\repsc$, and 
\item The probability that there exists some object $\good \in \goodset$ such that both $\price{\good}>0$ and $\sum_{\agent \in \agentset}\rvralloc{\good}{\agent} < \left(1-\epsfa\right) \supply{\good}$  is at most $\repsc$,
\end{enumerate}
where the randomness is over the random arrival permutation.
\end{lemma}

\proofst
Fix a triple $\triple = (\outputtypeset, \budgetfunc, \prices)$ and object $\good$. We first show that the probability that a sample $\sample$  is bad with respect to this triple and object is low. The sample is bad if $\sum_{\agent \in \sample}\ralloc{\good}{\agent}(\triple) \leq  \epss \supply{\good} \wedge \sum_{\agent \in \agentset}\ralloc{\good}{\agent}(\triple) > \left(1+\epsfa\right) \supply{\good}$. We bound the probability that this event happens; the probability is over the choice of the sample. 
For any triple $\triple$, set $\tempy{\agent} = \dfrac{\supply{\good}\ralloc{\good}{\agent}(\triple)}{\sum_{\agent \in \agentset}\ralloc{\good}{\agent}(\triple)}$.

\begin{align}
& \Pr \left[\sum_{\agent \in \sample}\ralloc{\good}{\agent}(\triple) \leq \epss \supply{\good} \wedge \sum_{\agent \in \agentset}\ralloc{\good}{\agent}(\triple) > \left(1+\epsfa\right) \supply{\good}\right] \notag \\
&\leq \Pr \left[\sum_{\agent \in \sample}\tempy{\agent} \leq  \epss \supply{\good} \wedge \sum_{\agent \in \agentset}\tempy{\agent} = \left(1+\epsfa\right) \supply{\good}\right]\notag\\
&\leq \Pr \left[\sum_{\agent \in \sample}\tempy{\agent} \leq \epss \supply{\good} \bigg|  \sum_{\agent \in \agentset}\tempy{\agent} = \left(1+\epsfa\right) \supply{\good}\right]\notag\\
&\leq  \Pr \left[  \left|\sum_{\agent \in \sample}\tempy{\agent}  - \epss \expect{\sum_{\agent \in \agentset}\tempy{\agent} } \right| \geq \epsfa \epss \supply{\good}\right] \notag\\
&\leq 2 \exp \left(-\frac{(\epsf \epss \supply{\good})^2}{64 \epss \supply{\good} +\epsf \epss \supply{\good} } \right) \label{feelTheBern}\\
&\leq 2 \exp \left(-\frac{\epsf^2 \epss \supply{\good}}{65} \right), \notag\\
& \leq \frac{1}{6\ngoods \nagents \cdot (\epss \nagents)^{\numtypes}}. \notag
\end{align}
Inequality~\eqref{feelTheBern} is obtained using Theorem~\ref{thm:Bern}, noting that  $\sum_{\agent=1}^\nagents (\tempy{\agent}-\frac{1}{\nagents}\sum_{\altagent = 1}^{\nagents}{\tempy{\altagent}})^2 \leq 2\supply{\good}$; hence $\sigma^2_\nagents \leq \frac{\supply{\good}}{\nagents}$. The last inequality is due to Assumption~\ref{ass1}.

We now take a union bound over all distinct triples and objects. Each triple $\triple = (\outputtypeset, \budgetfunc, \prices)$ is uniquely determined by $\outputtypeset$. Therefore it suffices to bound the possible values of $\outputtypeset$. As the sample size is $\epss \nagents$, each type can appear in the sample at most $\epss \nagents$ times; hence there  are at most $(\epss \nagents)^{\numtypes}$ possible values of $\outputtypeset$, and therefore of $\triple$.  While this is a loose upper bound, it is asymptotically tight if there are no restrictions on the type space. Taking a union bound over the objects and possible values of $\triple$ gives the first result. The proof for the second result is similar, only we have to bound the probability that $\sum_{\agent \in \sample}\ralloc{\good}{\agent}(\triple) \geq  \epss \supply{\good} \wedge \sum_{\agent \in \agentset}\ralloc{\good}{\agent}(\triple) < \left(1-\epsfa\right) \supply{\good}$ for objects $\good$ such that $\price{\good} > 0$. As the bound of  Theorem~\ref{thm:Bern} is symmetrical, the proof is virtually identical and omitted.  
\proofen

\subsection{Step 2: Bounding the deterministic allocation generated by the triple \texorpdfstring{$\triple$}{q}.}

\label{d2}
\begin{lemma}\label{lemma:det}
For any $\epsilons$, if Assumption~\ref{ass1} holds, then the (deterministic) allocation  and prices generated  by the triple $\triple$ satisfy the following: 
\begin{enumerate}
\item The probability that there exists some object $\good \in \goodset$ and $\knum \in [\epsn\nagents,\nagents]$ such that  $\sum_{\agent = 1}^{\knum} \rvalloc{\good}{\agent} > \left(1+\epsfc\right)\dfrac{\knum\supply{\good}}{\nagents}$  is at most $\repsa$ and 
\item  The probability that there exists some object $\good \in \goodset$ and $\knum \in [\epsn\nagents,\nagents]$ such that both $\price{\good}>0$ and  $\sum_{\agent = \realsample}^{\knum} \rvalloc{\good}{\agent} < (1-\epsfd)\dfrac{\knum\supply{\good}}{\nagents}$  is at most $\repsa$,
\end{enumerate}
where the probability is taken over the  random arrival permutation and the budget realization.
\end{lemma}

\proofst
For  any triple $\triple = (\outputtypeset, \budgetfunc, \prices)$ and object $\good$, denote the event  $\sum_{\agent \in \agentset}\ralloc{\good}{\agent}(\triple) >  \left(1+\epsfa\right)\supply{\good}$ by $\eventy$, 
 the event $\sum_{\agent = 1}^{\nagents} \rvalloc{\good}{\agent}(\triple) > (1+\epsfb) \supply{\good}$ by $\eventtwo$, and  the event  that there exists some $\knum \in [\epsn\nagents,\nagents]$ such that $\sum_{\agent = 1}^{\knum} \rvalloc{\good}{\agent}(\triple) > \left(1+\epsfc\right)\frac{\knum\supply{\good}}{\nagents}$ by $\eventone$. We first bound the probability of $\eventtwo$. 
\begin{align}
\Pr\left[\eventtwo\right] &\leq \Pr\left[\eventtwo |\neg \eventy \right] \Pr\left[ \neg \eventy \right] + \Pr\left[  \eventy \right] \notag\\ 
& \leq \Pr\left[\sum_{\agent \in \agentset}\rvalloc{\good}{\agent}(\triple) - \sum_{\agent \in \agentset}\ralloc{\good}{\agent}(\triple)> \epsft\supply{\good}\right]\label{reason1} + \repsc\\
&= \Pr\left[\sum_{\agent \in \agentset}\rvalloc{\good}{\agent}(\triple) - \expect{\sum_{\agent \in \agentset}\rvalloc{\good}{\agent}(\triple)}> \epsft\supply{\good}\right] +\repsc\notag \\
&\leq  \exp\left(-\frac{\epsf^2\supply{\good}}{32}\right) +\repsc \label{reason3}\\
&\leq \repsb,  \notag
\end{align}
where~\eqref{reason3} is from Theorem~\ref{Chernoff}, using the  fact that conditioned on $ \neg \eventy$, $\expect{\sum_{\agent = 1}^{\Nn} \rvalloc{\good}{\agent}} \leq (1+\epsfb)\supply{\good}$, and the last inequality is due to Assumption~\ref{ass1}.

We can now bound the probability of $\eventone$ using the bound on $\eventtwo$ and Theorem~\ref{Hoeffding}:

\begin{align}\Pr\left[\eventone\right] &\leq \Pr\left[\eventone |\neg \eventtwo \right] \Pr\left[ \neg \eventtwo \right] + \Pr\left[  \eventtwo \right] \notag\\ 
& \leq  \Pr\left[\max_{\knum \in [\epsn\nagents,\nagents]}\sum_{\agent = 1}^{\knum} \rvalloc{\good}{\agent}(\triple) > \left(1+\epsfc\right)\dfrac{\knum\supply{\good}}{\nagents}\bigg| \sum_{\agent = 1}^{\nagents} \rvalloc{\good}{\agent}(\triple) \leq  \left(1+\epsfb\right)\supply{\good}\right ] +\repsb\notag\\
& \leq  \Pr\left[\max_{\knum \in [\epsn\nagents,\nagents]} \left\{\sum_{\agent = 1}^{\knum} \rvalloc{\good}{\agent}(\triple) - \frac{\knum}{\nagents}\sum_{\agent = 1}^{\nagents} \rvalloc{\good}{\agent}(\triple) \right\}>\epsfa\dfrac{\knum\supply{\good}}{\nagents}\right ]+\repsb\notag\\ 
&\leq \exp{\left(\frac{- 2\realsample\epsf^2{\supply{\good}}^2}{16\nagents^2}\right)}+\repsb\label{eq:serfling}\\
&\leq \repsa,\notag
\end{align}
where Inequality~\eqref{eq:serfling} is due to Theorem~\ref{Hoeffding} and the last inequality is due to Assumption~\ref{ass1}.

The proof that conditioned on $\price{\good}>0$, $\Pr\left[\sum_{\agent = \realsample}^{\nagents}\rvalloc{\good}{\agent}<(1-\epsfd)\supply{\good}\right] \leq \repsa$ is similar and omitted.
\proofen

\subsection{Step 3: Bounding the deterministic allocation of the~\ocam}

\label{d3}

\begin{lemma}\label{lemma:detmech}
For any $\epsilons$, if Assumption~\ref{ass1} holds, the output of the~\ocam,  $(\optallocs,\optbudgets,\optprices)$ satisfies the following: 
\begin{enumerate}
\item The probability that there exists some object $\good \in \goodset$ and $\knum \in [\epsn\nagents,\nagents]$ such that  $\sum_{\iter = 1}^{\agentdef} \optgalloc{\good}{\iter} > (1+\epsfd)\dfrac{\knum\supply{\good}}{\nagents}$  is at most $\repsa$ and 
\item  The probability that there exists some object $\good \in \goodset$ and $\knum \in [\epsn\nagents,\nagents]$ such that both $\price{\good}>0$ and  $\sum_{\iter = 1}^{\agentdef} \optgalloc{\good}{\iter} < (1-\epsfd)\dfrac{\knum\supply{\good}}{\nagents}$  is at most $\repsa$,
\end{enumerate}
where the probability is taken over the  random arrival permutation and the budget realization.
\end{lemma}

\proofst
 The probability that $\exists \agentdef \in [\realsample,\nagents], \good \in \goodset$ such that $\sum_{\iter = 1}^{\agentdef} \optgalloc{\good}{\iter}  \geq  (1+\approxeps)\dfrac{\agentdef\supply{\good}}{\nagents}$  can be bounded as follows:
 
 \begin{align*}   \Pr\left[ \sum_{\iter = 1}^{\agentdef} \optgalloc{\good}{\iter}  >(1+\approxeps)\dfrac{\agentdef\supply{\good}}{\nagents}\right]  
 \leq   \Pr\left[ \sum_{\iter = 1}^{\agentdef} \rvalloc{\good}{\agent}
> \left(1+\epsfc\right)\dfrac{\agentdef\supply{\good}}{\nagents}\right]
 \leq \repsa,
\end{align*}

where the first inequality is because 
\begin{enumerate}
\item $\sum_{\iter = 1}^{\knum} \optgalloc{\good}{\iter} \leq  \sum_{\iter = 1}^{\realsample} \optgalloc{\good}{\iter}+  \sum_{\iter = 1}^{\agentdef} \rvalloc{\good}{\agent}$, and 
\item $\sum_{\iter = 1}^{\realsample} \optgalloc{\good}{\iter}  \leq \epss\supply{\good} \leq   \frac{\epsf\epsn}{4}\supply{\good} \leq \epsfa\frac{\agentdef\supply{\good}}{\nagents},$ 
\end{enumerate}
and the second inequality is from Lemma~\ref{lemma:det}.
Similarly, if $\price{\good}>0$, we have 

 \begin{align*}   \Pr\left[ \sum_{\iter = 1}^{\agentdef} \optgalloc{\good}{\iter}  <  (1-\approxeps)\dfrac{\agentdef\supply{\good}}{\nagents}\right]  
 \leq   \Pr\left[ \sum_{\iter = \realsample}^{\agentdef} \rvalloc{\good}{\agent}
<  \left(1-\epsfd\right)\dfrac{\agentdef\supply{\good}}{\nagents}\right]
 \leq \repsa,
\end{align*}
where the first inequality is because $\sum_{\iter = 1}^{\knum} \optgalloc{\good}{\iter} = \sum_{\iter = 1}^{\realsample} \optgalloc{\good}{\iter}+  \sum_{\iter = \realsample}^{\agentdef} \rvalloc{\good}{\agent}$, and the second inequality is from Lemma~\ref{lemma:det}.
\proofen

\subsection{Step 4: Putting it all together}
\label{d4}

\proof{Proof of Theorem~\ref{thm1}.}
Parts~\ref{thm1a} and~\ref{thm1b}  of Theorem~\ref{thm1} are proven in  Lemma~\ref{lemmathm1}. From the construction of the allocations in the \ocam, the output $(\optallocs,\optbudgets,\optprices)$ trivially satisfies Parts~\ref{ACEEI1}  and~\ref{ACEEI2}  of Definition~\ref{def:aceeb} (for all $k \in [1,n]$).   Lemma~\ref{lemma:detmech} shows that  $(\optallocs,\optbudgets,\optprices)$ satisfies Parts~\ref{ACEEI3} and~\ref{ACEEI4} of Definition~\ref{def:aceeb} with probability at least $\frac{1}{n}$.
\proofen

\clearpage

\section{Pseudocode for the OCAM}

\begin{algorithm}[htpb]

\medskip

	\SetAlgoNoLine
	\KwIn{A set $\goodsset$ of goods  with capacity  $\supplyvector{}$, $\epsilons$, $\nagents$ agents arrive online.}
	\KwOut{Allocation $\allocv{}$, budgets $\budgets$ and prices $\prices$.}
    Set $\epss = \frac{\epsf \epsn}{4}$\;
    \For{each  agent $\agent \in [\realsample]$}
        {Agent $\agent$ reports their type $\ttype_{\agent}$\;   
    
        Set $\allocv{\agent} = \max_{\succ_{\agent}}\{\bundle{}: \bundle{} \in \choices_{\agent} \text{ and } \sum_{\ell=1}^{\agent} \allocv{\ell} \leq \epss \supplyvector\}$\;
    
         Arbitrarily set $\budget{\agent} \in [1-\epsb,1]$\;
        }
 
    Define the economy $\econ'=([\realsample],\iden,M, \epss\supplyvector{}, (\choices_\agent)_{\agent\in[\realsample]}, (\succ_\agent)_{\agent \in [\realsample]})$, where $\iden$ is the identity permutation\;
    
    Compute an $\epsb$-\eceei\  for  $\econ'$:  $\bundlev, \randombudget{1}, \ldots, \randombudget{\realsample},  \prices$ \;

    Set $ \sampletypeset = \cup_{\agent \in \{1, \ldots, \realsample\}} \ttype_{\agent}$\; 
        
    Define $\budgetfunc: \sampletypeset \rightarrow \{\randombudget{1}, \ldots, \randombudget{\realsample}\}$ as follows:  $\budgetfunc(\ttype) = \randombudget{\agent}$, where 
    $\ttype_{\agent} = \ttype $ for some $\agent \in [\realsample]$\; 

    \For{each  agent $\agent \in \{ \realsample+1, \ldots, \nagents\}$} {Agent $\agent$ reports their type $\ttype_{\agent}$\;
    \eIf{$\ttype_\agent \in \sampletypeset$}{Draw $\budget{\agent} \sim \budgetfunc(\ttype_\agent)$\;}
    {Arbitrarily set $\budget{\agent} \in [1-\epsb,1]$\;} 
    Set $\allocv{\agent} = \max_{\succ_{\agent}}\{\bundle{}: \bundle{} \in \choices_{\agent} \text{ and } \prices\cdot \bundle{} \leq \budget{\agent}\}$\;} 

	\caption{\ocam}
	\label{alg:one}
\end{algorithm}

\end{document}